\documentclass[preprints,article,accept,moreauthors,pdftex]{mdpi} 
\usepackage{subfigure}
\makeatletter
\renewcommand{\@thesubfigure}{\normalsize(\textbf{\alph{subfigure}})}
\makeatother

\firstpage{1} 
\pubvolume{xx}
\issuenum{1}
\articlenumber{5}
\pubyear{2019}
\copyrightyear{2019}
\history{Dated: 28 July 2020}

\Title{Hybrid Quantum-Classical Neural Network for Calculating Ground State Energies of Molecules}

\Author{Rongxin Xia $^{1,2,3}$ and Sabre Kais $^{1,2,3}$*}

\AuthorNames{Rongxin Xia and Sabre Kais}

\address{%
$^{1}$ \quad Department of Chemistry, Purdue~University, West Lafayette, IN 47906, USA; xiar@purdue.edu\\
$^{2}$ \quad Department of Physics and Astronomy, Purdue~University, West Lafayette, IN 47906, USA\\
$^{2}$ \quad Birck Nanotechnology Center, Purdue~University, West Lafayette, IN 47906, USA\\}
\corres{\hangafter=1 \hangindent=1.05em \hspace{-0.82em}Correspondence: kais@purdue.edu}

\abstract{We present a hybrid quantum-classical neural network that can be trained to perform electronic structure calculation and generate potential energy curves of simple molecules. The~method is based on the combination of parameterized quantum circuits and measurements. With~unsupervised training, the neural network can generate electronic potential energy curves based on training at certain bond lengths. To demonstrate the power of the proposed new method, we present the results of using the quantum-classical hybrid neural network to calculate ground state potential energy curves of simple molecules such as H$_2$, LiH, and BeH$_2$. The results are very accurate and the approach could potentially be used to generate complex molecular potential energy surfaces.}


\begin{document}


\section{Introduction}
Quantum computing has shown its great potential in advancing quantum chemistry research~\cite{kais-book}. Many quantum algorithms have been proposed to solve quantum chemistry problems ~\cite{cao2018quantum,xia2017electronic, bian2019quantum}, such~as the Phase Estimation Algorithm;~\cite{aspuru2005simulated,wang2008quantum,whitfield2011simulation,daskin2018direct} to calculate eigenstate energies of simple molecules; the~Variational Quantum Eigensolver (VQE)~\cite{o2016scalable, peruzzo2014variational, kandala2017hardware} to solve electronic structure problems; quantum algorithms for open quantum dynamics~\cite{hu2019quantum}; and benchmark calculations for two-electron molecules conducted on quantum computers~\cite{smart2020efficient}. Using quantum computing techniques to perform machine learning tasks~\cite{biamonte2017quantum} has also received much attention recently including quantum data \mbox{classification~\cite{rebentrost2014quantum, lloyd2013quantum},}  quantum generative learning~\cite{dallaire2018quantum,lloyd2018quantum},  and quantum neural network approximating nonlinear functions~\cite{mitarai2018quantum}. So~far, applying the various quantum machine learning techniques to quantum chemistry is a natural extension~\cite{ romero2017quantum,xia2018quantum}. However, previous studies focused solely on quantum circuits with only a few nonlinear operations, which are introduced by data encoding~\cite{mitarai2018quantum,schuld2019quantum} or repeated measurements until success~\cite{cao2017quantum}. Moreover, recently Sim et. al~\cite{sim2019expressibility} shows increasing the number of layers of the parameterized quantum circuit (PQC) would reach saturation and may not improve the performance when the number of layers is large enough. Furthermore, the nonlinearity is the most important part for the classical neural network~\cite{cybenko1989approximation} which makes neural networks able to produce complex results~\cite{cao2017quantum,hopfield1982neural,hinton2006reducing}. Therefore, quantum machine learning should not solely focus on PQC and nonlinear operations are needed for the quantum neural network.

To solve this problem, here we introduce a new hybrid quantum-classical neural network,  by combining quantum computing and classical computing  with  measurements between the parameterized quantum circuits. In this paper, we first give a detailed description of the whole structure of the hybrid quantum-classical neural network. We then present numerical simulations by using the  new hybrid quantum-classical neural network to calculate ground state energies of different molecular  systems. The calculated ground state energies are very accurate, which demonstrate the potential of the proposed hybrid quantum-classical neural network to generate potential energy surfaces. 
 
\section{Results}

\begin{figure}[H]
    \centering
    \includegraphics[width=5in]{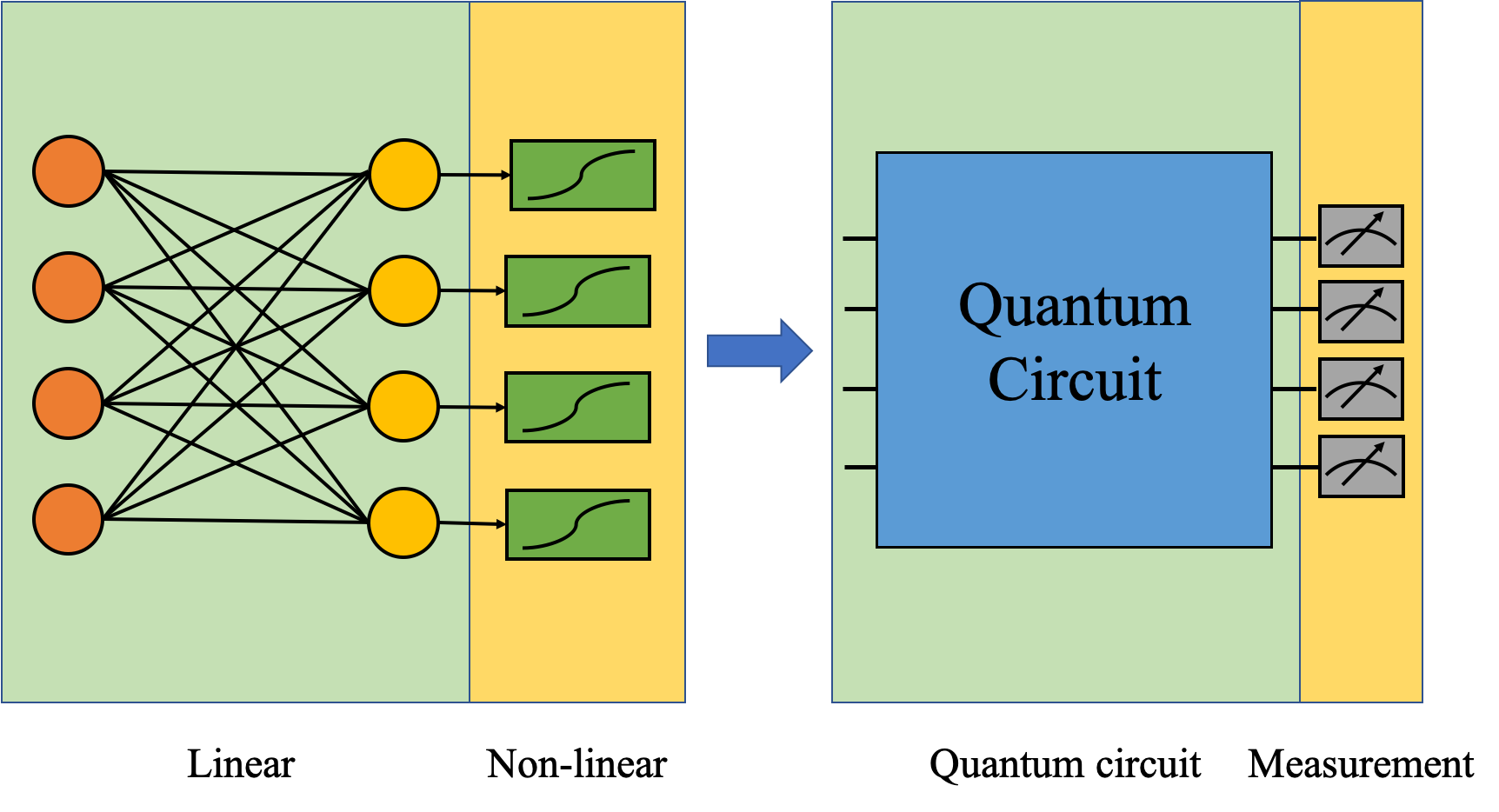}
    \caption{In the proposed quantum-classical hybrid neural network, the linear part in the classical neural network is replaced by the quantum circuits and the nonlinear part is replaced by  measurements.}
    \label{quantum-classical}
\end{figure}

We propose a new structure of quantum-classical hybrid neural network by connecting the quantum part (quantum layer) with the classical part (classical layer). For a classical neural network, each artificial neuron is normally constructed by linear connected layers, with nonlinear activation functions connected at the end,  as shown in the left part of Figure~\ref{quantum-classical}. In this work, we replace the linear part by the quantum circuit as shown in the right side of Figure~\ref{quantum-classical}  to take advantage of possible speedup in quantum computation. In the meantime, we use expectation values of operators by measurements, which are nonlinear operations, to serve as the activation function. In this neural network set-up, the~quantum circuit can be viewed as the quantum layer and the expectation values by measurements can be viewed as the classical layer. The input data is first encoded into quantum states and calculated by the quantum layer. The outputs are extracted as the expectation values by  measurements.  The~two steps can be repeated several times to construct a hybrid multi-layer neural network. In our construction, the quantum layer is enabled by parameterized quantum circuits (PQC)~\cite{du2018expressive}. We will give details about the hybrid quantum-classical neural network in the following~sections.

\subsection{Quantum Layer}
The quantum layer is enabled by a parameterized quantum circuit consisting of parameterized quantum gates, which allows the PQC to be optimized by adjusting the parameters to approximate wanted results. PQC has been widely used in many areas of quantum computing and quantum machine learning, such as in VQE~\cite{o2016scalable, peruzzo2014variational, kandala2017hardware}, quantum autoencoder~\cite{romero2017quantum}, and quantum generative learning~\cite{dallaire2018quantum}. In the following section, we will provide details of the quantum layer including encoding classical data into quantum circuits and parameterized quantum circuits.

\subsubsection{Data Encoding}
To implement the quantum layer, the first step is to encode the input classical data into a quantum state. Variational encoding~\cite{schuld2019quantum} has been proposed to reduce the depth of quantum circuits and has been widely used in many quantum machine learning techniques~\cite{mitarai2018quantum, schuld2019quantum,romero2019variational,levine2018bridging}. Variational encoding is used to prepare a set of quantum gates with parameters generated by the input data and then initialize the state from the basic state with all qubits as $|0\rangle$ with these gates. For an array of data $\{a_0, a_1, ... a_{n-1}\}$, an example of variational encoding to encode $n$ qubits is to prepare the gate $G$ as

\begin{equation}
    G = \otimes_{i=0}^{n-1} g_i(f_i(a_i))
\end{equation}
where $g_i$ is a set of single qubit quantum gates on qubits $i$ and $f_i$ is a classical function to encode $a_i$ as the parameter of $g_i$. The encoded state would be $G|0\rangle^{\otimes n}$.  One simple example is given in our numerical simulations: we take the bond length, $a$, as the encoding data for each qubit. We choose $f_i$ as the identity function and $g_i$ as $R_yH$, where $R_y$ is the rotation-$y$ gate and $H$ is the Hadamard gate. Thus, the encoded quantum state would be $(\otimes_{i=0}^{n-1}R_y(a)H)|0\rangle^{\otimes n}$.

In most variational encoding the depth of the circuit needed to encode the data would be $O(1)$~\cite{romero2019variational} for that the number of quantum gates to initialize the quantum state is fixed, which makes variational encoding more suitable for Noisy Intermediate-Scale Quantum (NISQ) devices~\cite{preskill2018quantum}. Furthermore, recently it has been shown how the variational encoding may help to introduce nonlinearity features in quantum circuits~\cite{schuld2019quantum, huggins2018towards}. Variational encoding can only be implemented at the beginning of the quantum circuit, but connections between multiple PQC also need to be nonlinear. To enable nonlinear connections, we introduce measurements as connections between multiple PQC. In the numerical simulations, we will be using the variational encoding to perform the simulation and discuss implementing the quantum circuits on NISQ device.

\subsubsection{Parameterized Quantum Circuit}
A parameterized quantum circuit, also known as a variational quantum circuit~\cite{peruzzo2014variational, du2018expressive}, is a quantum circuit consisting of parameterized gates with fixed depth. This is the main part of the quantum layer to perform the calculation.  The parameterized quantum circuit consists of one-qubit gates as well as $CNOT$. Some more complicated gates may also be used in PQC which can be decomposed into one qubit gates and $CNOT$~\cite{nielsen2002quantum}. In general, an $n$ qubits PQC can be written as

\begin{equation}
    U(\vec{\theta})|\psi\rangle = (\prod_{i=1}^m U_i)|\psi\rangle
\end{equation}
where $U(\vec{\theta})$ is the set of universal gates and $m$ is the number of quantum gates. $\vec{\theta}$ is the set of parameters $\{\theta_0, \theta_1....\theta_{k-1}\}$, where $k$ is the total number of parameters and  $|\psi\rangle$ is the encoded quantum state after data encoding. For each unitary gate $U_i$, it may be a quantum gate which does not require a parameter or a quantum gate which takes parameters. Examples of the unitary gate taking parameters are rotational gates, $R_x(\theta)$, $R_y(\theta)$, and $R_z(\theta)$, which are given by

\begin{equation}
    R_x(\theta) = e^{-i\frac{\theta}{2}\sigma_x}  \qquad  R_y(\theta) = e^{-i\frac{\theta}{2}\sigma_y} \qquad  R_z(\theta) = e^{-i\frac{\theta}{2}\sigma_z} 
\end{equation}
where $\sigma_x$, $\sigma_y$, and $\sigma_z$ are Pauli matrices. The operation of $U$ can be modified by changing parameters $\vec{\theta}$. Thus, the output state can be optimized to approximate the wanted state by changing parameters $\vec{\theta}$. By optimizing the parameters used in $U(\vec{\theta})$, PQC approximates the  wanted quantum states. 

\subsection{Classical Layer}
The classical layer in our construction of the quantum-classical hybrid neural network is to serve as the activation function connecting different quantum layers. To achieve nonlinearity, the classical layer is enabled by measurements---expectation values of operators on each qubit of the PQC, for example, $\langle \sigma_z^i \rangle$ of each qubit $i$ as the classical layer, which would also serve as nonlinear operations. Expectation values of operators can save complexity because quantum tomography is exponentially hard. Though the expectation values of operators may lose some information compared to quantum tomography, some work used expectation values of operators as connections between quantum computation and classical computation and showed great success~\cite{mari2019transfer}, which indicates expectation values of operators are capable of extracting useful information from quantum circuits.

\subsection{Numerical Simulations}
To demonstrate the power of the proposed quantum-classical hybrid neural network, we present results for calculating the  ground state energies of  simple  molecular systems:  H$_2$, LiH, and BeH$_2$. The~inputs for the unsupervised learning are  bond lengths and the outputs are the ground state energies.  The whole procedure consists of  first  training  the neural network with some bond lengths and then testing the neural network with other bond lengths to generate the whole potential energy~curve.

\subsubsection{Constructions of the Quantum Layer}
\begin{figure}[H]
    \centering
    \includegraphics[width=4in]{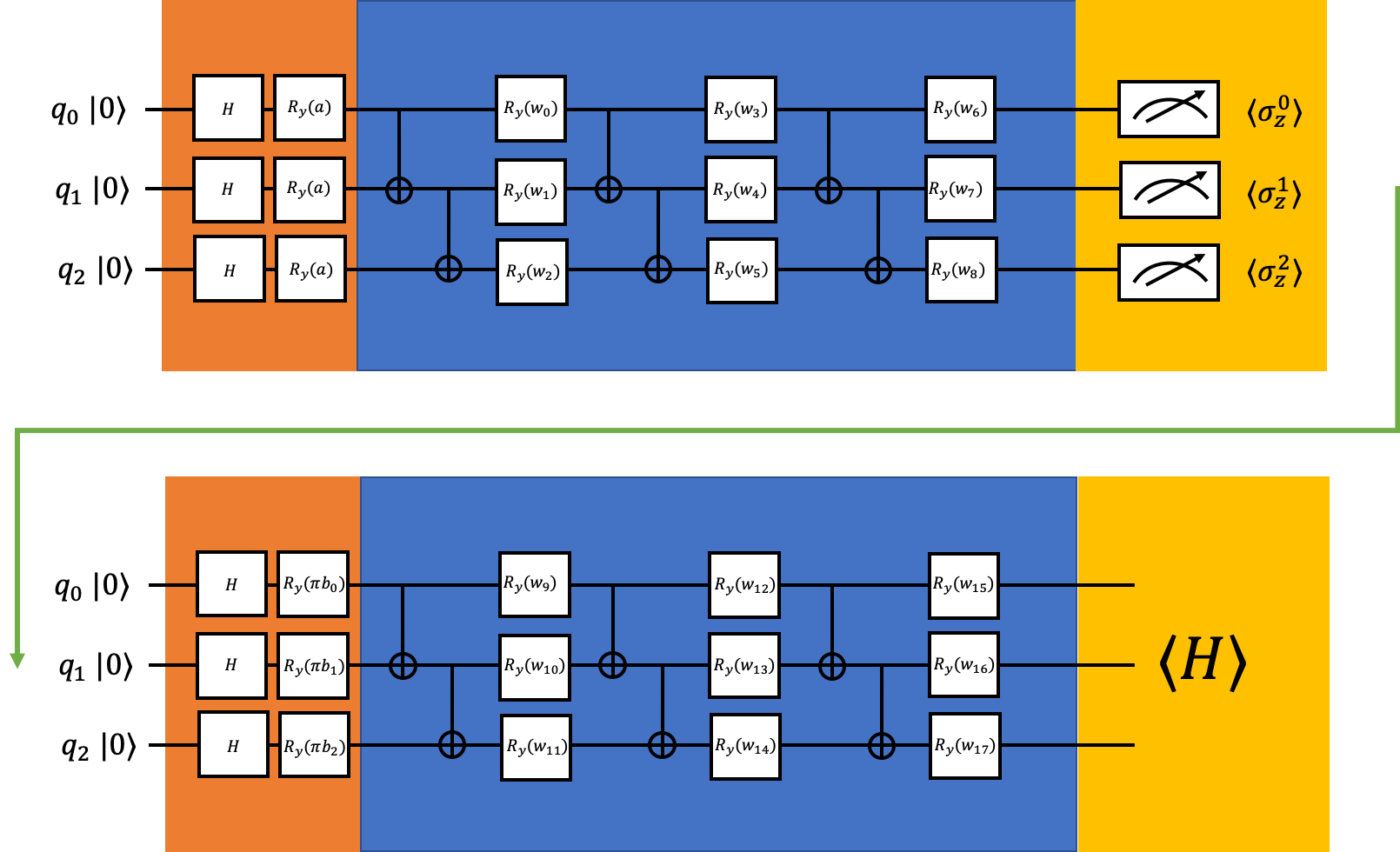}
    \caption{The example constructions of the proposed hybrid quantum-classical neural network for 3 qubits (odd qubits number). The orange parts are the data encoding, the blue parts are parameterized quantum circuits, and the yellow parts are measurements. The first measurements serve as nonlinear operations connecting two PQC.  $a$ is the input bond length, $b$s are the expectation values of $\sigma_z$, and $w$s are adjustable parameters.}
    \label{qcircuit_1}
\end{figure}

\begin{figure}[H]
    \centering
    \includegraphics[width=4in]{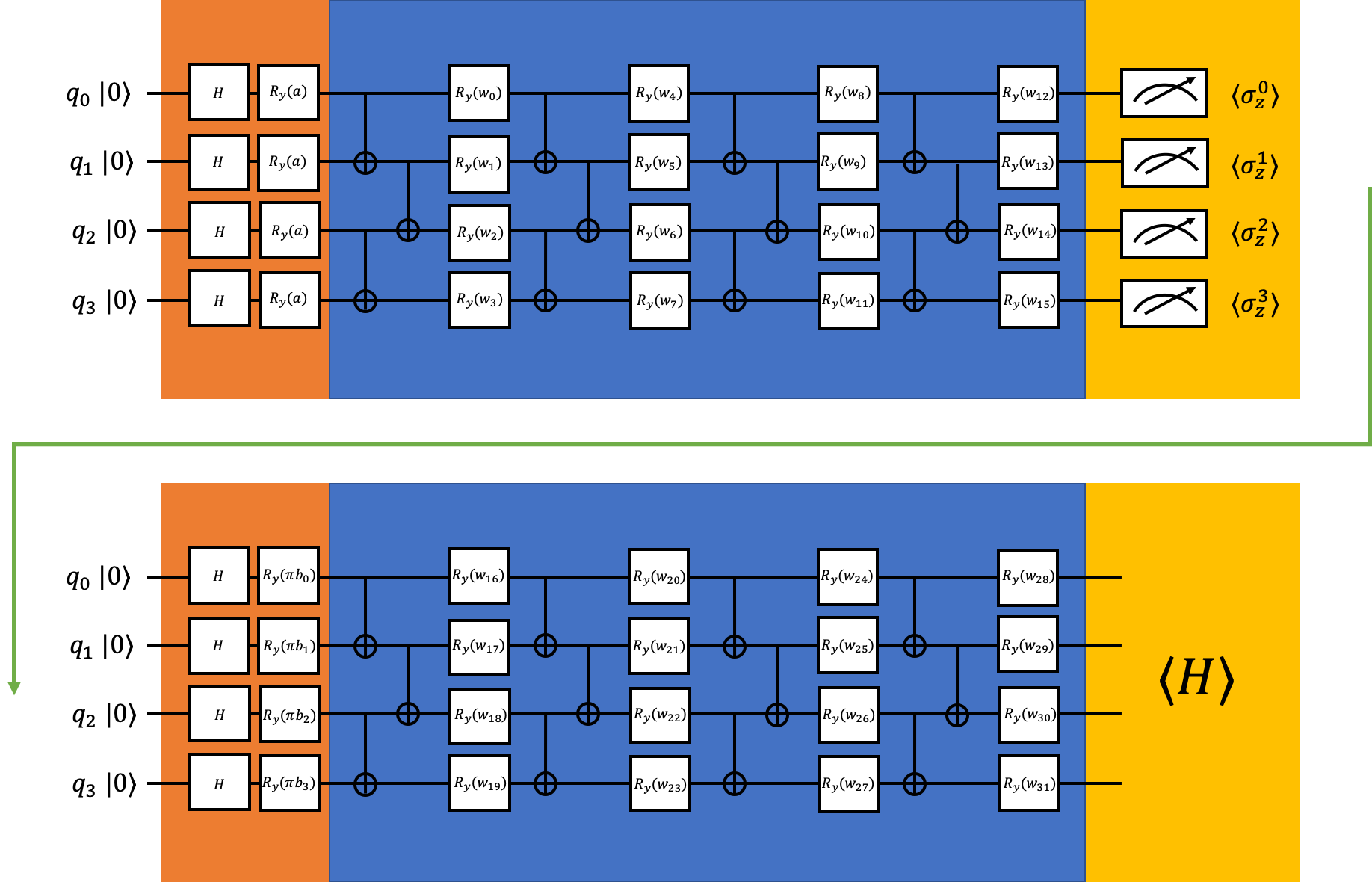}
    \caption{The example constructions we use for the 4 qubits H$_2$ calculation (even qubits number). The~orange parts are the data encoding, the blue parts are parameterized quantum circuits, and the yellow parts are measurements. The first measurements serve as nonlinear operations connecting two PQC.  $a$ is the input bond length, $b$s are the expectation values of $\sigma_z$, and $w$s are adjustable parameters.}
    \label{qcircuit}
\end{figure}

The quantum layer consists of two parts: the variational encoding part and PQC part.  We~choose to use the variational encoding to decrease the depth of the quantum circuit so that it can be implemented on NISQ devices. The construction of the quantum layer follows  ~\cite{romero2019variational,mari2019transfer}. The input state is initialized as $(\otimes_{i=0}^{n-1}R_y(a)H)|0\rangle^{\otimes n}$, where $a$ is the bond length, $H$ is the Hadamard gate, and $R_y$ is the rotation-$y$ gate. We only have one bond length while the number of qubits of the PQC is $n$; we decided to follow the variational encoding in~\cite{romero2019variational} to encode each qubit with same value. The number of qubits $n$ is equal to the number of qubits of the corresponding Hamiltonian. The quantum computation part is to use a simple PQC consisting of $R_y$ and $CNOT$ gates, which can be written as
\begingroup\makeatletter\def\f@size{9}\check@mathfonts
\def\maketag@@@#1{\hbox{\m@th\large\normalfont#1}}%
\begin{equation}
\begin{aligned}
    &\prod_{j=0}^{n-1}(\otimes_{i=0}^{n-1}R_y(w_{i+n\times j}))(CNOT_{n-3,n-2}...CNOT_{3,4}CNOT_{1,2})(CNOT_{n-2,n-1}...CNOT_{2,3}CNOT_{0,1})\ n \text{ is even}\\
    &\prod_{j=0}^{n-1}(\otimes_{i=0}^{n-1}R_y(w_{i+n\times j}))(CNOT_{n-2,n-1}...CNOT_{3,4}CNOT_{1,2})(CNOT_{n-3,n-2}...CNOT_{2,3}CNOT_{0,1})\ n \text{ is odd}
\end{aligned}
\end{equation}
\endgroup
where $w$ are adjustable parameters, $R_y$ represents rotation-$y$ gate, and $CNOT_{m,n}$ represents $CNOT$ gate with $m$ as the control qubit and $n$ is the target qubit. To achieve better entanglement of the qubits before appending nonlinear operations, the $n$ qubits PQC has $n$ repeated layers in our simulation. By~optimizing the parameters, the general PQC tries to approximate arbitrary states so that it can be used for different specific molecules.  The construction of the PQC for three qubits is illustrated in the blue part of Figure~\ref{qcircuit_1}, and the construction of the PQC for four qubits is illustrated in the blue part of Figure~\ref{qcircuit}.

\subsubsection{Constructions of the Classical Layer}
The classical layer is enabled by expectation values of the operators. In our numerical simulations, we are using $\langle \sigma_z^i \rangle$ for qubit $i$ as the classical layer. The outputs from the classical layer will be encoded into another quantum layer. The second quantum layer is the same as the first one except for the data encoding part it would be $\otimes_{i=0}^{n-1}R_y(b_i\pi)H$, where $b_i$ is the measured expectation value from qubit $i$.  We multiply each $b_i$ with $\pi$ when encoding to change the range of the encoding data from $[-1,1]$ to $[-\pi,\pi]$~\cite{stoudenmire2016supervised}. The construction of our proposed hybrid quantum-classical neural network is illustrated in Figure~\ref{qcircuit}.

\subsubsection{Cost Function}
The cost function is defined as

\begin{equation}
    f = \sum_{j}\langle\phi_j|H_j|\phi_j\rangle
\end{equation}
where $j$ represents the $j_{th}$ input bond length of the training bond lengths. $|\phi_j\rangle$ is the final state of the proposed hybrid quantum-classical neural network with the input as the $j_{th}$ input bond length and $H_j$ is the Hamiltonian corresponding to the $j_{th}$ input bond length. The idea of the cost function is similar to VQE: by optimizing the parameters, the expectation energy of $|\phi_j\rangle$ is minimized to approximate the ground state energy.   The evaluation of the Hamiltonian can be done by techniques in ~\cite{kandala2017hardware}. The~Hamiltomian can be written as the sum of tensor products of Pauli matrices $H = \sum_i c_iP_i$, where $c_i$ is the coefficient and $P_i$ is the tensor product of Pauli matrices. Instead of evaluating the whole Hamiltonian, we can evaluate each term of the Hamiltonian and the expectation of the Hamiltonian can be obtained by  $\langle H\rangle = \sum_ic_i \langle P_i\rangle$, which does not need quantum tomography or take exponential complexity. The~whole training procedure is done by taking a set of bond lengths and corresponding Hamiltonian and minimizing the cost function as equation (5). After the training, we test the model with other bond~lengths.

\subsubsection{Simulation Results}
The Hamiltonian of the molecule systems can be derived by transforming the  corresponding second quantization Hamiltonian into sum of tensor products of Pauli matrices. For H$_2$, we use the Jordan--Wigner transformation~\cite{fradkin1989jordan} to  get a 4-qubit Hamiltonian. We decided to apply the complete active space (CAS) approach~\cite{roos1980complete,romero2018strategies}, which divides the orbitals into inactive orbitals such as always occupied low energy orbitals and always unoccupied high energy orbitals, and active orbitals, to~reduce the number of qubits of LiH and BeH$_2$ Hamiltonian~\cite{kandala2017hardware,bian2019quantum} and the reduced Hamiltonian is only of the active orbitals. For LiH, we assume the first two lowest energy spin orbitals are always occupied and use the binary code transformation~\cite{steudtner2017lowering} considering spin symmetry to save two qubits. We get an 8-qubit LiH Hamiltonian. For BeH$_2$, we assume the first two lowest energy spin orbitals are always occupied and the first two highest energy spin orbitals are never occupied, and use the binary code transformation~\cite{steudtner2017lowering} considering spin symmetry to save two qubits. We get an 8-qubit BeH$_2$~Hamiltonian. 

\begin{figure}
    \centering 
  \subfigure[]{ 
    \includegraphics[width=4.5in]{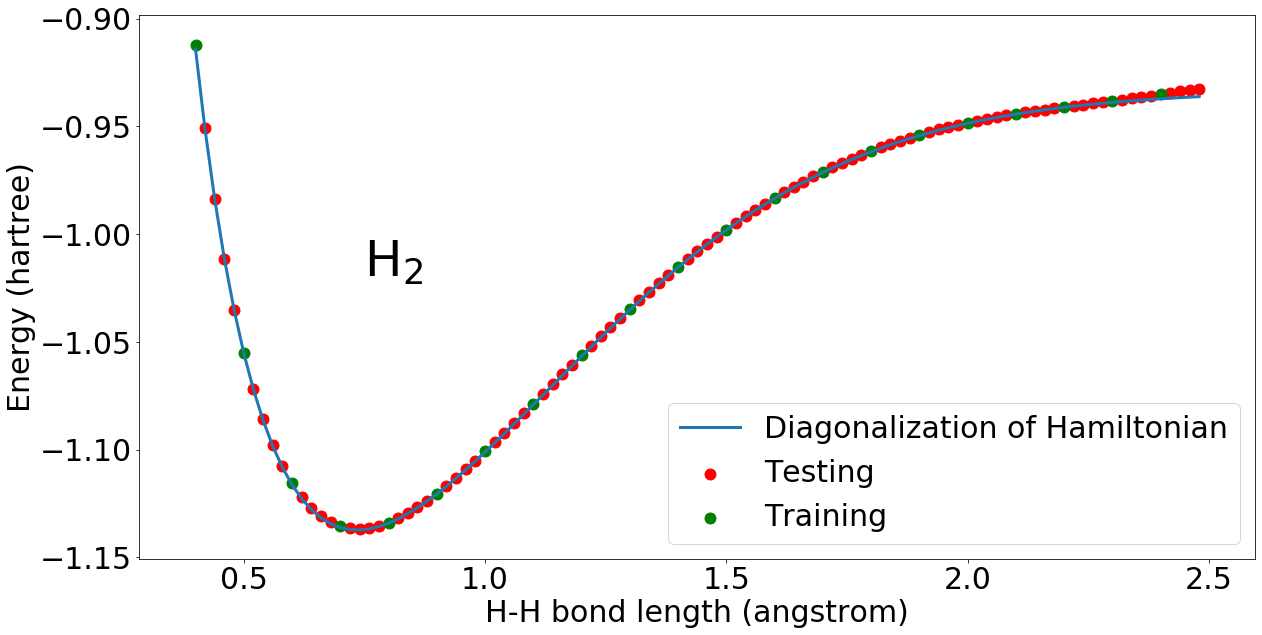}} 
    \subfigure[]{ 
    \includegraphics[width=4.5in]{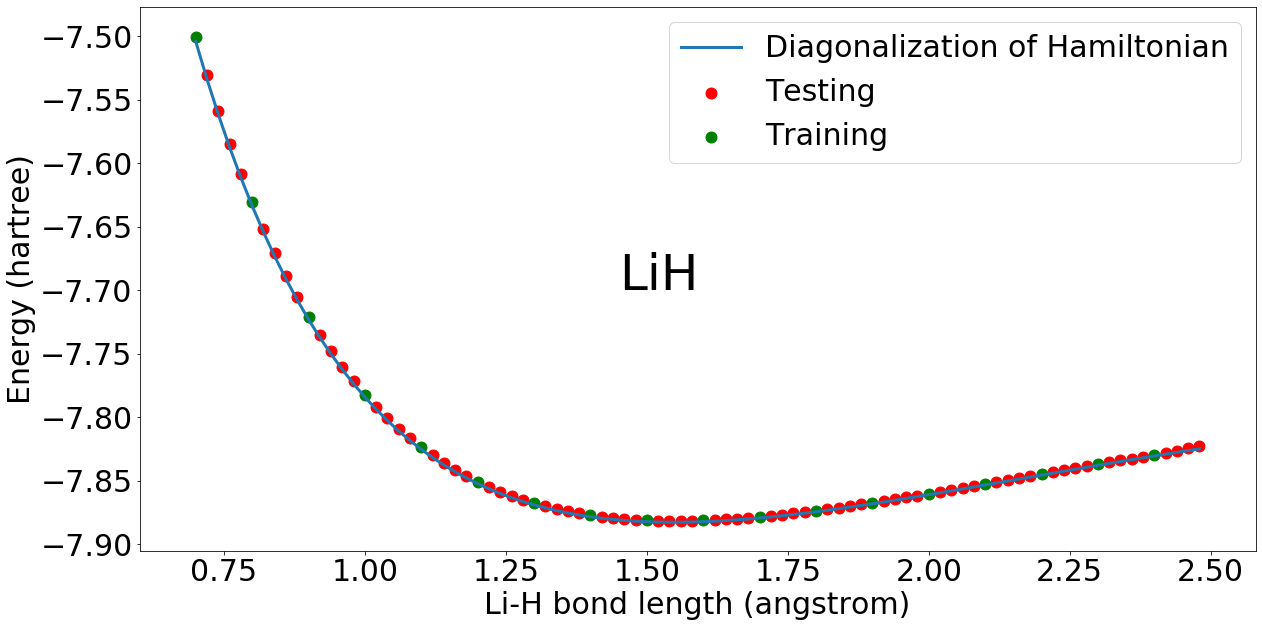}} 
    \subfigure[]{ 
    \includegraphics[width=4.5in]{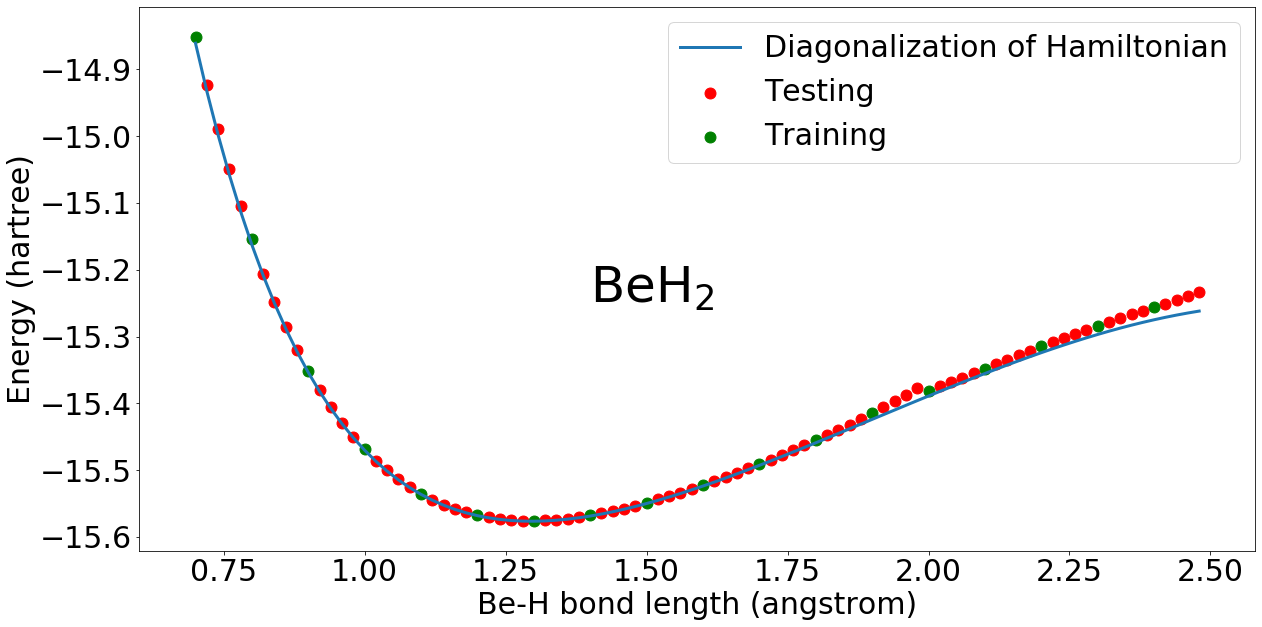}} 
    \caption{Ground state energies of H$_2$, LiH, and BeH$_2$ calculated by the proposed hybrid quantum- classical neural network.~(\textbf{a}) Ground state energies of H$_2$ calculated by the proposed hybrid quantum-classical neural network; (\textbf{b}) Ground state energies of LiH calculated by the proposed hybrid quantum-classical neural network; (\textbf{c}) Ground state energies of BeH$_2$ calculated by the proposed hybrid quantum-classical neural network.}
    \label{results_1}
\end{figure}

\begin{figure}
    \centering 
  \subfigure[]{ 
    \includegraphics[width=4in]{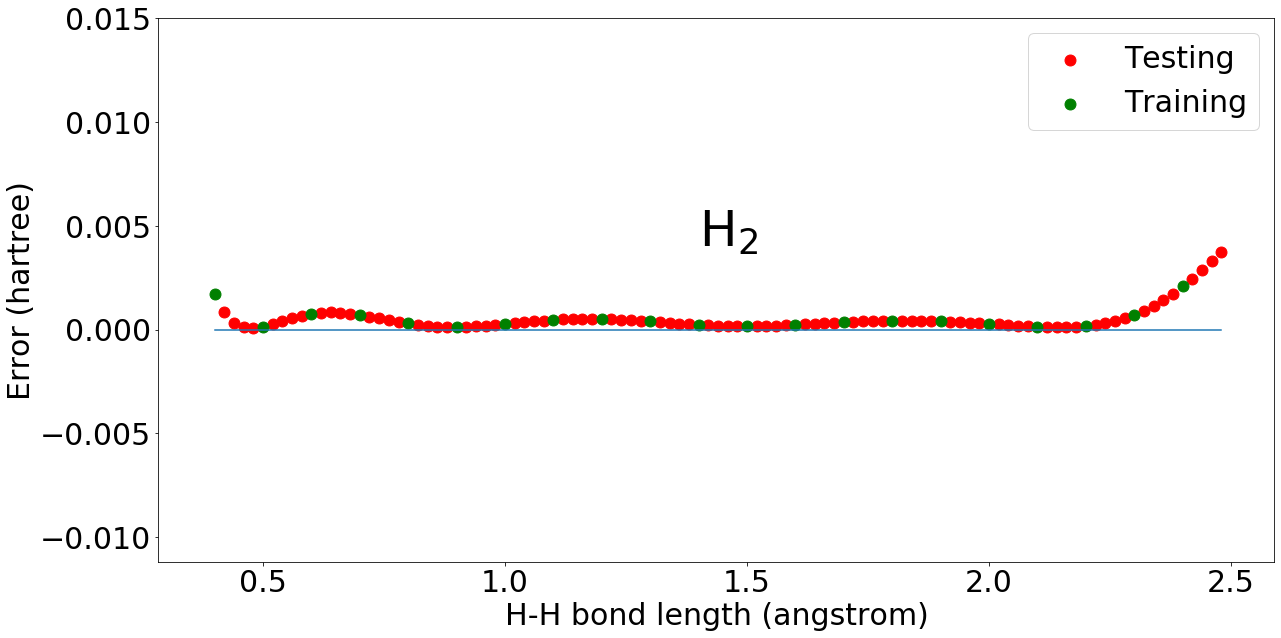}} 
    \subfigure[]{ 
    \includegraphics[width=4in]{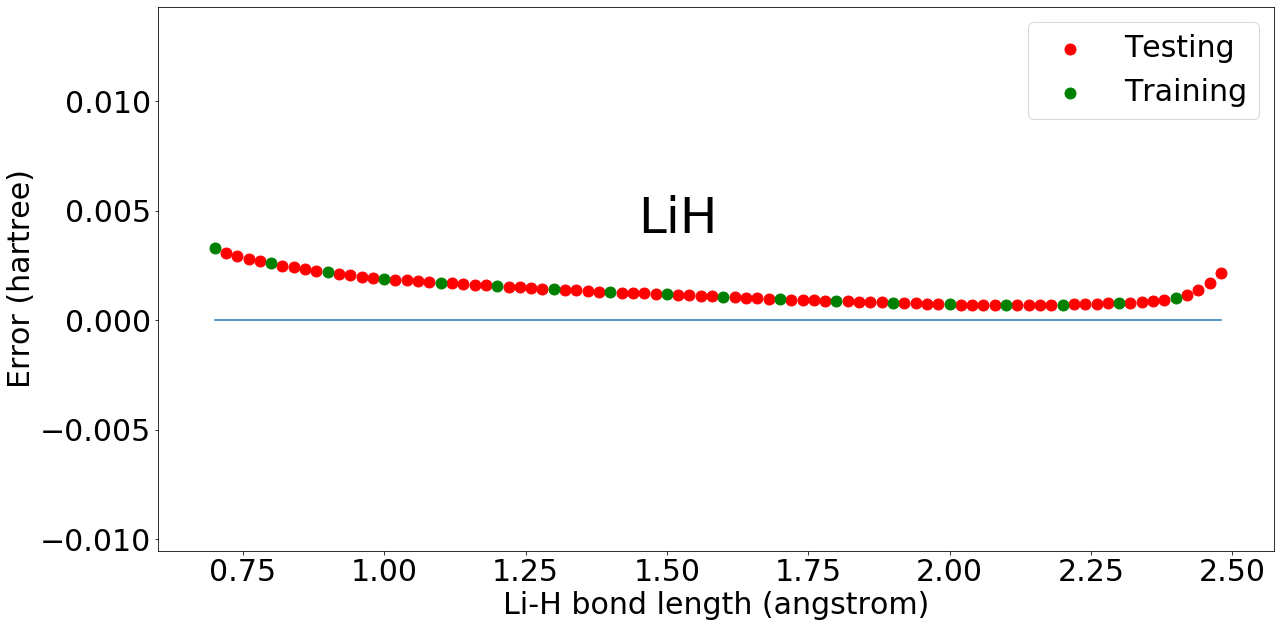}} 
    \subfigure[]{ 
    \includegraphics[width=4in]{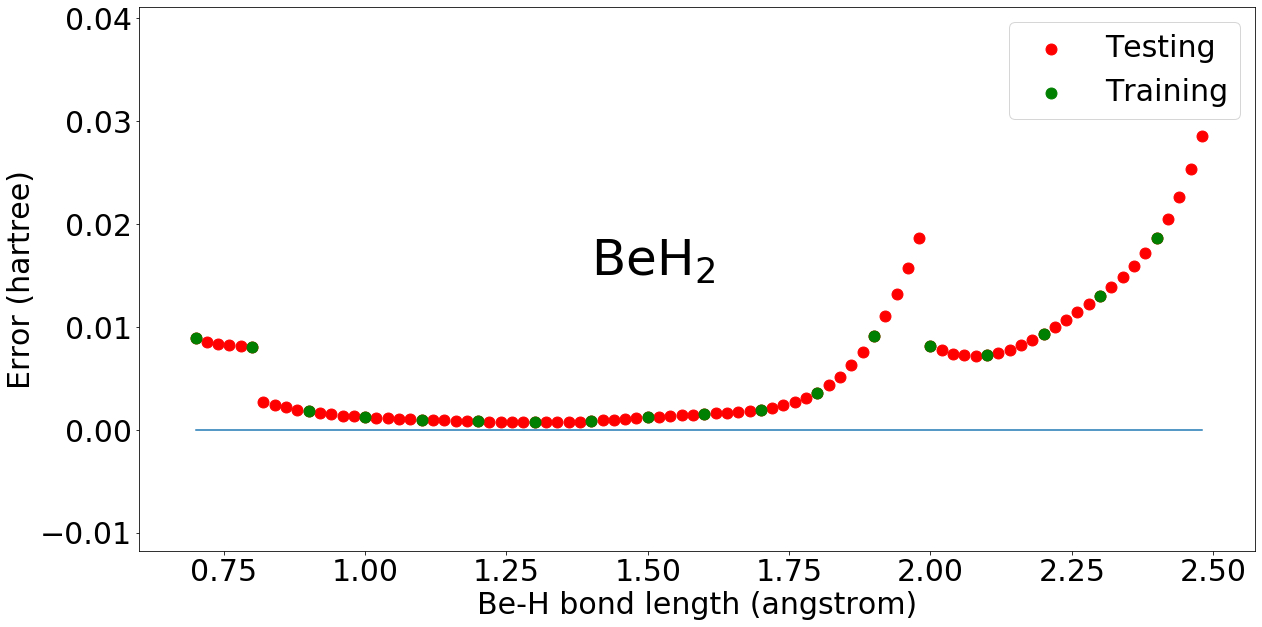}} 
    \caption{Errors of ground state energies of H$_2$, LiH, and BeH$_2$ calculated by the proposed hybrid quantum-classical neural network. (\textbf{a}) Errors of ground state energies of H$_2$ calculated by the proposed hybrid quantum-classical neural network; (\textbf{b}) Errors of ground state energies of LiH calculated by the proposed hybrid quantum-classical neural network; (\textbf{c}) Errors of ground state energies of BeH$_2$ calculated by the proposed hybrid quantum-classical neural network.}
    \label{results_2}
\end{figure}

In the simulation, H$_2$ used four qubits and 32 parameters. LiH and BeH$_2$ both used eight qubits and 128 parameters. The gate and parameter complexity of the proposed hybrid quantum-classical neural network in this simulation is $O(n^2)$, where $n$ is the number of qubits of the Hamiltonian. Here, we present the results using our proposed hybrid quantum-classical neural network for ground state energies of H$_2$, LiH, and BeH$_2$ in Figures \ref{results_1} and \ref{results_2}. We can see from these figures that the training data points converge very close to the diagonalization results without pre-known ground state information of the transformed Hamiltonian in Pauli matrices format. Furthermore, after training, by inputting the other bond lengths we can also get good approximating ground state energies with optimized parameters. BeH$_2$ has some deviation when the bond length is large, which may be solved by improving the parameterized quantum circuit. For example, the work in~\cite{sim2019expressibility}, which discusses expressibility and entangling capability of parameterized quantum circuits for hybrid quantum‐classical algorithms, shows that increasing the depth of PQC will increase the expressibility and different constructions of PQC have also different expressibility.

\begin{figure}
    \centering
    \includegraphics[width=6in]{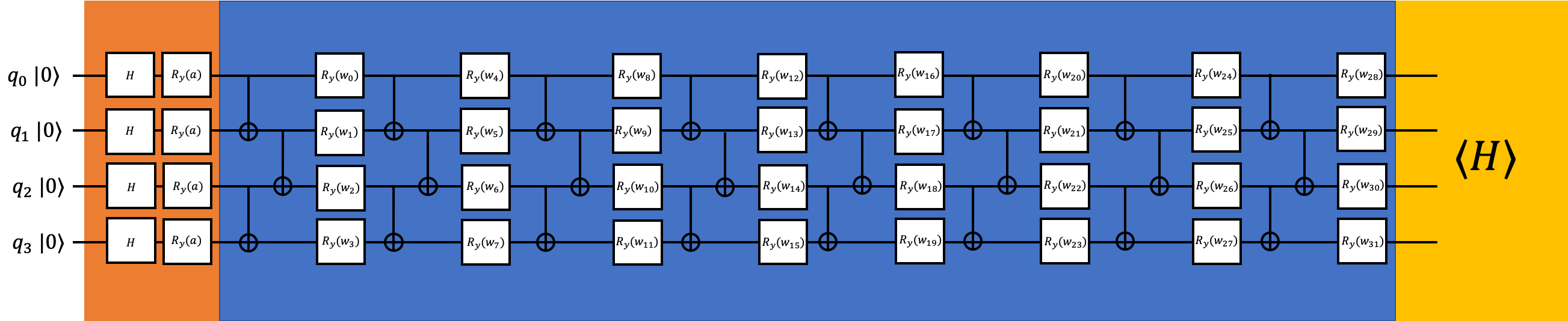}
    \caption{Constructions of the quantum neural network removing intermediate measurements for H$_2$.}
    \label{qcircuit_li}
\end{figure}

Furthermore, to show that the intermediate nonlinear measurements improve the performance, we present the comparison of the results of our proposed hybrid quantum-classical neural network and quantum neural network removing intermediate measurements. The setting of the quantum neural network removing intermediate measurements is illustrated in Figure~\ref{qcircuit_li}.

\begin{figure}
  \centering 
  \subfigure[]{ 
    \includegraphics[width=4in]{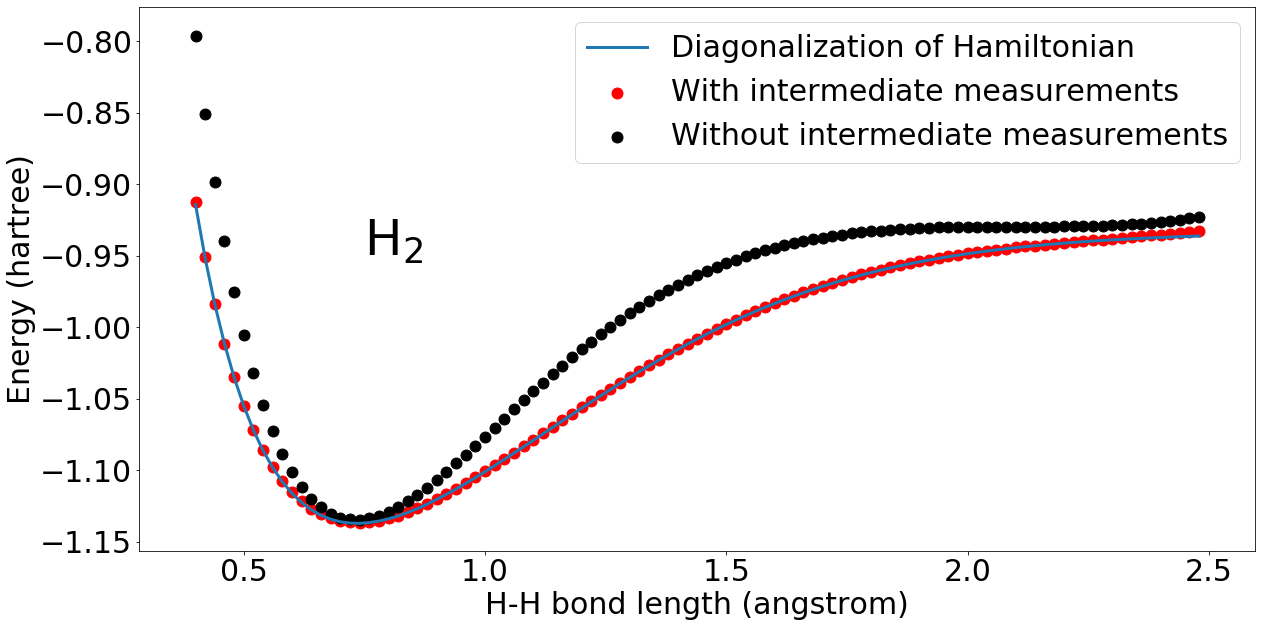}} 
  \hspace{0in} 
  \subfigure[]{ 
    \includegraphics[width=4in]{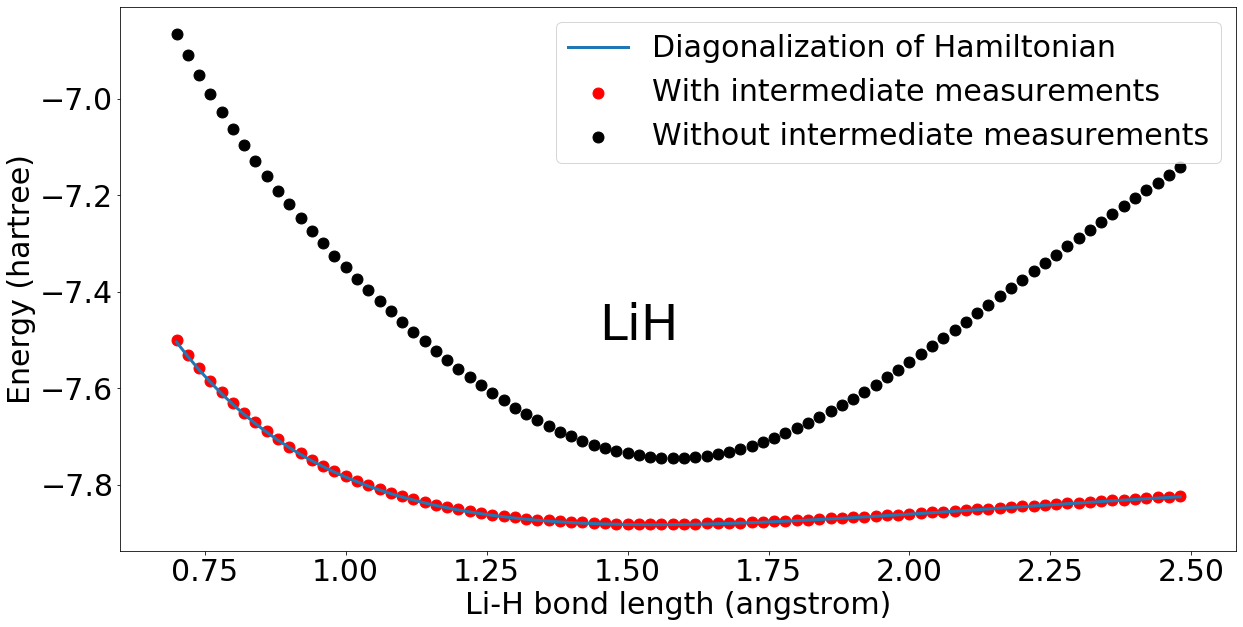}}
    \hspace{0in}
  \subfigure[]{ 
    \includegraphics[width=4in]{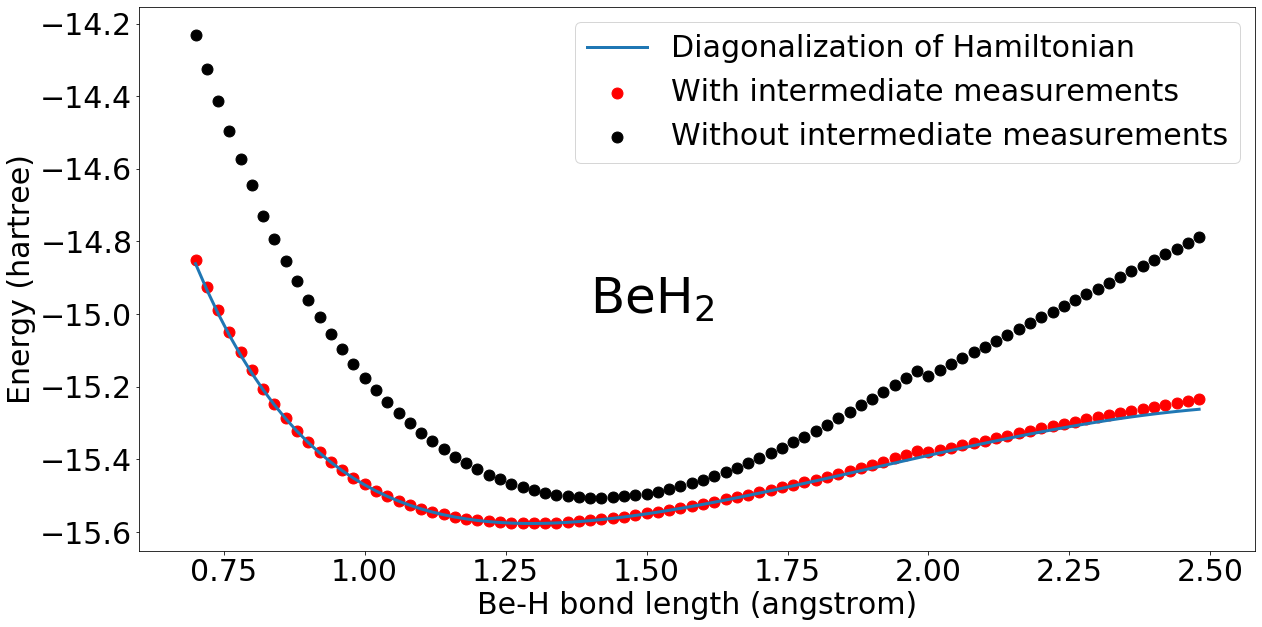}} 
    \caption{Results of H$_2$, LiH, and BeH$_2$ by the proposed hybrid quantum-classical neural network and the quantum neural network removing intermediate measurements. With intermediate measurements represents the results by our proposed hybrid quantum-classical neural network. Without intermediate measurements represents the quantum neural network removing the intermediate measurements. Both~are trained with same set of bond lengths as in Figure~\ref{results_1} and same parameter initialization. (\textbf{a})~Ground state energies of H$_2$ calculated by the proposed hybrid quantum-classical neural network and the quantum neural network removing intermediate measurements; (\textbf{b}) Ground state energies of LiH calculated by the proposed hybrid quantum-classical neural network and the quantum neural network removing intermediate measurements; (\textbf{c}) Ground state energies of BeH$_2$ calculated by the proposed hybrid quantum-classical neural network and the quantum neural network removing intermediate measurements.}
    \label{li}
\end{figure}

In Figure~\ref{li}, we present the comparison of the results of our proposed hybrid quantum-classical neural network and quantum neural network removing intermediate measurements. The proposed hybrid quantum-classical neural network and quantum neural network, removing intermediate measurements, are trained with same set of bond lengths as in Figure~\ref{results_1}. We can see without the intermediate nonlinear measurements, the quantum neural network can only achieve bad results. However, by adding the intermediate nonlinear measurements, the results converge closely to the diagonalization results.

The parameters of the proposed hybrid quantum-classical neural network and quantum neural network removing intermediate measurements, are initialized from a Gaussian distribution with standard deviation as 0.1 and mean as 0. Because different initialization of parameters will result in different starting of the optimization and may lead to different final results, to eliminate the effects of parameter initializations, here we present the quantitative comparison of the two constructions with four different parameter initialization from same Gaussian distribution with different random seeds. All are trained with the same set of the training bond lengths as in Figure~\ref{results_1}. In the Table \ref{tab1}, we can see that our proposed quantum neural network performs better than the quantum neural network without intermediate measurements. Our simulation results show that adding intermediate nonlinear measurements would help to improve the expressibility of the PQC. Furthermore, adding intermediate measurements would also decrease the circuit depth which makes it more suitable for current NISQ~devices.

\vspace{-6pt}

\begin{table}[H]
\centering
\caption{Results for the proposed hybrid quantum-classical neural network and the quantum neural network removing intermediate measurements. With intermediate measurements represents the proposed hybrid quantum-classical neural network. Without intermediate measurements represents the quantum neural network removing intermediate measurements. $\sum_{training} Error$ represents the sum of the error of the calculated ground state energies on the training set. $\sum_{testing} Error$ represents the sum of the error of the calculated ground state energies on the testing set. Each result is calculated by 4 different parameter initialization and presented as means and standard deviations. It can be seen that adding intermediate measurements to introduce nonlinear options would help to improve the~performance.}
\label{tab1}
\begin{tabular}{lll} \toprule
\bf{Constructions} & \boldmath{$\sum_{training} Error$} & \boldmath{$\sum_{testing} Error$}   \\ \midrule
With intermediate measurements (H$_2$) & $0.0271\pm 0.0246$ & $0.1178\pm 0.1061$\\
Without intermediate measurements (H$_2$)& $0.6296\pm 0.0151$ & $2.2755\pm 0.0677$\\
With intermediate measurements (LiH) & $0.0287\pm 0.0038$ & $0.1178\pm 0.0190$\\
Without intermediate measurements (LiH)& $4.7638\pm 1.4444$ & $19.1479\pm 5.7715$\\
With intermediate measurements (BeH$_2$) & $0.1253\pm 0.0552$ & $0.5613\pm 0.2483$\\
Without intermediate measurements (BeH$_2$)& $3.7280\pm 0.6497$ & $14.8440\pm 2.3747$\\
 \bottomrule
\end{tabular}
\end{table}

\section{Materials and Methods}
Orbital integrals in the second quantization Hamiltonian are calculated by STO-3G minimal basis using PySCF~\cite{PYSCF} and the transformation is done by OpenFermion~\cite{mcclean2017openfermion}. The simulation is done by Qiskit~\cite{Qiskit}. The tensor production orders in OpenFermion and Qiskit are opposite. For a $n$ qubits, the tensor production order in OpenFermion is $q_0\otimes q_1...\otimes q_{n-1}$, while the tensor production order in Qiskit is $q_{n-1}\otimes q_{n-2}...\otimes q_0$. We decided to follow the tensor production order in OpenFermion. In~simulation, we treat the qubit indexed in Qiskit reversely. For $n$ qubits, the qubit indexed as $q_{0}$ in Qiskit is treated as $q_{n-1}$, the qubit indexed as $q_{1}$ in Qiskit is treated as $q_{n-2}$, etc. By doing this, we change the tensor production order in Qiskit same as OpenFermion. The optimization is performed by the Broyden--Fletcher--Goldfarb--Shanno algorithm~\cite{nocedal2006numerical} with maximum 500 iterations and gradient norm tolerance to stop as $10^{-5}$. In the simulation, the expectation of the operator is simulated by matrix production of the operator matrix and the Hamiltonian can be treated as a single operator. To~save the simulation time, instead of evaluating each $\langle P_i\rangle$ to get $\langle H\rangle = \sum_ic_i \langle P_i\rangle$, we treat H as a single operator and only evaluate once. 
\section{Conclusions}
In this work, we proposed a new hybrid quantum-classical neural network by combing PQC and measurements to achieve nonlinear operations in quantum computing. We have shown that the proposed hybrid quantum-classical neural network can be trained to obtain the electronic energies at certain bond lengths   and then generate the whole potential energy curve. The results of H$_2$, LiH, and~BeH$_2$ are very accurate and demonstrate the power of the proposed hybrid quantum-classical neural network. 

Furthermore, we show that the  intermediate nonlinear measurements are very important in comparison with quantum neural network removing the intermediate measurements. The intermediate nonlinear measurements can reduce the circuit depth and are more suitable for NISQ devices. Although the method is used to generate one-dimensional potential energy curves, the approach is general and could be generalized to generate multidimensional potential energy surfaces, for example, changing the inputs from the bond lengths to multidimensional coordinates. This will be done in future work.   

\vspace{6pt} 



\authorcontributions{S.K. designed the research. R.X. performed the calculations. Both discussed the results and wrote the paper. All authors have read and agreed to the published version of the manuscript.}

\funding{We  acknowledge the financial support by the U.S. Department of Energy (Office of Basic Energy Sciences) under Award No. DE-SC0019215 and the Integrated Data Science Initiative Grant (IDSI F.90000303), Purdue University.}

\acknowledgments{The authors would like to thank Zixuan Hu for critical reading and useful discussions.}

\conflictsofinterest{The authors declare no conflicts of interest.} 

\abbreviations{The following abbreviations are used in this manuscript.\\

\noindent 
\begin{tabular}{@{}ll}
VQE & Variantional Quantum Eigensolver\\
PQC &  parameterized quatum circuit\\
NISQ & Noisy Intermediate-Scale Quantum\\
\end{tabular}}


\reftitle{References}





\end{document}